\documentclass[twocolumn,secnumarabic,amssymb, nobibnotes, aps, prl]{revtex4}
\usepackage{graphicx}
\usepackage{amsmath,dsfont}
\usepackage{overpic}

\newcommand{\beq}{\begin{equation}}
\newcommand{\eeq}{\end{equation}}
\newcommand{\beqa}{\begin{eqnarray}}
\newcommand{\eeqa}{\end{eqnarray}}

\begin{document}
\title{\bf Comment on ``Surprises in threshold antikaon-nucleon physics''}
\vskip 1.5true cm
\author{B. Borasoy$^{a}$}
%\email{borasoy@itkp.uni-bonn.de}
\author{R. Ni{\ss}ler$^{a}$}
%\email{rnissler@itkp.uni-bonn.de}
\author{W. Weise$^{b}$}
%\email{weise@ph.tum.de}
\affiliation{\it $^{a}$ Helmholtz-Institut f\"ur Strahlen- und Kernphysik (Theorie),
             Universit\"at Bonn, Nu{\ss}allee 14-16, D-53115 Bonn, Germany}
\affiliation{\it $^{b}$ Physik Department, Technische 
Universit\"at M\"unchen, D-85747 Garching, Germany}

\begin{abstract}
It has recently been claimed \cite{OPV} that the DEAR kaonic hydrogen data
can be reconciled with $K^- p$ scattering data in a chiral unitary
approach. In this comment we demonstrate that the proposed solution
in \cite{OPV} violates fundamental principles of scattering theory.
%\begin{center}
%\textbf{PACS:} 11.80.-m, 12.39.Fe, 13.75.Jz, 36.10.Gv  \quad\quad
%\textbf{Keywords:}  Chiral Lagrangians, coupled channels, unitarity.
%\end{center}
\end{abstract}

% 11.80.-m   Relativistic scattering theory 
% 11.80.Gw   Multichannel scattering
% 12.39.Fe   Chiral Lagrangians
% 13.75.Gx   Pion-baryon interactions
% 13.75.Jz   Kaon-baryon interactions
% 36.10.Gv   Mesonic atoms and molecules, hyperonic atoms and molecules

\maketitle

%%%%%%%%%%%%%%%%%%%%%%%%%%%%%%%%%%%%%%%%%%%%%%%%%%%%%%%%%%%%%%%%%%%%%%%%%%%%%%

Now that new accurate results for the strong-interaction
shift and width of kaonic hydrogen from the DEAR experiment are available \cite{DEAR},
there is renewed interest in an improved analysis of the $K^- p$ system.
In general,
non-perturbative coupled-channel techniques based on driving terms from the chiral SU(3) 
effective Lagrangian have proved successful in the strangeness $S=-1$ sector. 
However, a thorough investigation \cite{BNW} of low-energy $K^- p$ interactions within such approaches pointed to questions of consistency of the DEAR experiment
with previously measured sets of $K^- p$ scattering data. 
In contrast, it has been claimed very recently by Oller et al. \cite{OPV} that within a chiral unitary approach both the scattering and the DEAR data can be accommodated.
In the following we demonstrate that the proposed solution \cite{OPV} to this problem violates
basic principles of scattering theory and is rather an artifact of the model.

We have first reproduced the $A_4^+$ fit of \cite{OPV}. The $s$-wave interaction kernel
of the coupled-channels approach is derived from the Weinberg-Tomozawa term, contact 
interactions of next-to-leading chiral order as well as direct and crossed Born terms.
The $s$-wave contribution from the crossed Born term, however, leads to unphysical 
subthreshold cuts which are an artifact of the 
on-shell formalism used in the coupled-channels approach. These unphysical cuts would not be present in a full field theoretical calculation as discussed in some detail in \cite{BNW}.
The subthreshold singularities induce imaginary pieces in the interaction kernel 
and hence spoil exact unitarity of the approach.
As a test case we have explicitly checked the $S$-matrix element for 
$\pi^0 \Lambda \to \pi^0 \Lambda$, the channel with the lowest threshold. Its modulus deviates from 1
in the region below the first inelastic threshold ($\pi \Sigma$). Such unitarity violations can in fact be sizeable.

\begin{figure}[t]
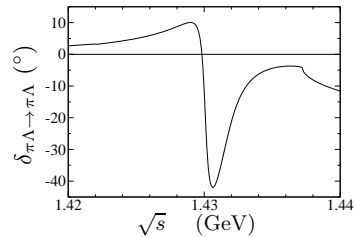

\centering
\begin{overpic}[width=0.23\textwidth]{PhSh_piLambda.eps}
\put(-10,15){\rotatebox{90}{\scalebox{1.0}{$\delta_{\pi \Lambda \to \pi \Lambda}$ ($^\circ$)}}}
\put(30,-6){\scalebox{0.9}{$\sqrt{s}$ \quad (GeV)}}
\end{overpic}
\caption{$\pi^0  \Lambda \to \pi^0 \Lambda$ phase shift for the fit $A_4^+$ of \cite{OPV}.}
\label{fig:phaseshifts}
\end{figure}

Nonetheless, we adopt the approach of \cite{OPV} in order to investigate their results
in more detail. A study of the analytical continuation of the $T$-matrix into the complex 
$\sqrt{s}$-plane reveals an isospin $I=1$ pole at $\sqrt{s} = (1431 + 1.3i)$ MeV, i.e.,
right at the $K^- p$ threshold and {\it above} the physical region.
Besides the fact that such a pronounced pole with a width of about 3 MeV on the real axis
is not seen empirically (e.g. in the $\pi \Sigma$ invariant mass spectrum),
it certainly violates fundamental principles of scattering theory. An unphysical 
$T$-matrix pole close to the physical region leads in its vicinity to a strong phase 
shift variation with energy as shown in Fig.~\ref{fig:phaseshifts}.
Such a steep decrease of a phase shift is in sharp contradiction to the Wigner condition which---based
on a general causality principle for any finite-range interaction---imposes a certain
bound on the rate at which the phase shift can change with energy \cite{Wig}.
The limit is given by the range of the underlying interaction, i.e.\ roughly
1\,fm in the case of strong interactions, whereas the steep fall-off in Fig.~\ref{fig:phaseshifts} would
require an interaction range of more than 300\,fm in order to fulfill the Wigner condition.
The appearance of the $I=1$ pole in the upper half plane therefore 
violates causality and the postulate of maximal analyticity in $S$-matrix theory.

We conclude by emphasizing that the proposed solution of \cite{OPV} and the artificial
pole arise for certain highly fine-tuned combinations of the effective 
Lagrangian parameters due to some
shortcomings of the applied coupled-channels approach, but should disappear
in a full field theoretical calculation. This is further substantiated by the observation 
that tiny variations in the parameter values of fit $A_4^+$ lead to sizeable changes in the
$K^- p$ scattering length, since the unphysical pole shifts slightly its position, 
bringing the fit into disagreement with the DEAR measurement.

We thank J.~A.~Oller for comparing the results of \cite{OPV} with ours.\\[-1cm]

\end{document}